\documentclass[twocolumn,showpacs,superscriptaddress]{revtex4}
\usepackage{amsmath,bm}
\usepackage{graphicx}
\usepackage{epsfig}
\begin{document}

\title{Suppression of high $p_T$ hadrons in Pb+Pb collisions at LHC}
\author{Subrata Pal}
\affiliation{Department of Nuclear and Atomic Physics, Tata Institute of
Fundamental Research, Homi Bhabha Road, Mumbai 400005, India} 
\author{Marcus Bleicher}
\affiliation{Institut f\"ur Theoretische Physik, Goethe-Universit\"at and 
Frankfurt Institute for Advanced Studies (FIAS), 60438 Frankfurt am Main, Germany}

\begin{abstract} 
Hadron production and their suppression in Pb+Pb collisions at LHC at a center-of-mass
energy of $\sqrt{s_{NN}} = 2.76$ TeV are studied within a multiphase transport (AMPT) 
model whose initial conditions are obtained from the recently updated HIJING 2.0 model. 
The centrality dependence of charged hadron multiplicity $dN_{\rm ch}/d\eta$ at midrapidity 
was found quite sensitive to the largely uncertain gluon shadowing parameter $s_g$ that 
determines the nuclear modification of the gluon distribution. We find final-state parton 
scatterings reduce considerably hadron yield at midrapidity and enforces a smaller gluon 
shadowing to be consistent with $dN_{\rm ch}/d\eta$ data at LHC. With such a constrained 
parton shadowing, charged hadron and neutral pion production over a wide transverse momenta 
range are investigated in AMPT. Relative to nucleon-nucleon collisions, the particle yield 
in central heavy ion collisions is suppressed due to parton energy loss. While the calculated 
magnitude and pattern of suppression is found consistent with that measured in Au+Au collisions 
at $\sqrt{s_{NN}} = 0.2$ TeV at RHIC, at the LHC energy the suppression is overpredicted which 
may imply the medium formed at LHC is less opaque than expected from simple RHIC extrapolations. 
Reduction of the QCD coupling constant $\alpha_s$ by $\sim 30\%$ in the higher temperature plasma 
formed at LHC as compared to that at RHIC was found to reproduce the measured suppression at LHC.  
\end{abstract}

\maketitle

Heavy ion collisions at the Relativistic Heavy Ion Collider (RHIC) \cite{BRAHMS,PHOBOS,STAR,PHENIX} 
and recently at the Large hadron Collider (LHC) \cite{ALICE,CMS} have revealed a new state of
matter comprising of strongly interacting quarks and gluons (sQGP) \cite{DIS,Shuryak}. 
Primary evidence of this is provided by the observed suppression of high 
transverse momenta single hadron spectra \cite{PHENIXsup,STARsup} in central collisions 
relative to both peripheral and nucleon-nucleon collision. The suppression has been
established as due to energy loss by the propagating partons in the plasma primarily by 
radiative gluon emission \cite{GLV,WS}. 
Since the parton scatterings occur at the early stage of the evolution in nuclear collisions, 
study of energy loss can probe the sQGP phase of the matter. In fact, the magnitude of energy 
loss is predicted to be strongly dependent on the parton density of the medium
which reappears as soft hadrons \cite{WS,PalPratt}. 

In addition to the final state parton energy loss, the jet quenching at moderate and high $p_T$ 
is also influenced by initial spatial distribution of partons, collective flow, and to the 
unknown nuclear shadowing of the parton distribution. As the matter created in Pb+Pb collisions 
at $\sqrt{s_{NN}} = 2.76$ TeV is at about twice the density and probes parton distribution at 
a smaller momentum fraction $x$ than at RHIC, analysis of the recent data for bulk hadron 
production \cite{ALICEnp,CMSnp} and high-$p_T$ hadron suppression at LHC \cite{ALICEsup,CMSsup} 
may provide crucial insight into the nuclear medium effects of parton shadowing and energy loss
in the hot and dense QCD matter.

While perturbative quantum chromodynamics (pQCD) can address only hard scatterings, 
formation of strongly coupled near perfect fluid as well as abundant soft particle production 
suggest a highly nonperturbative physics which is not yet well-established within QCD. 
Consequently models based on (non-)ideal hydrodynamics \cite{Paul,Song,Chen}, 
transport calculations \cite{UrQMD,AMPT,BAMPS}, and transport/hydrodynamics 
hybrid models \cite{Petersen} have been developed.
It was recently demonstrated in the HIJING 2.0 model \cite{HIJ2} that the larger 
uncertainties of the shadowing effects at RHIC \cite{Li} can be constrained 
from comparison of the measured charged particle density at midrapidity for the most 
central Pb+Pb collision at LHC. On the other hand, collision centrality dependence of 
bulk hadron observables should reflect the relative contribution to particle production from hard 
and soft processes. Thus a precise estimate of nuclear shadowing and detailed study 
of medium effects on particle production from soft to the hard scattering regime relies on 
systematic inclusion of various stages of dynamical evolution of matter.

A MultiPhase Transport (AMPT) model \cite{AMPT} which combines the initial particle distribution 
from HIJING model \cite{HIJING} with subsequent parton-parton elastic scatterings via the
ZPC parton cascade model and final hadron 
transport via ART allows a systematic study of hadron production and jet quenching.
In this letter we shall investigate bulk charged particle production and jet suppression
within the AMPT model modified to include the updated HIJING 2.0 version.
In absence of control d+Pb data essential to calibrate the
nuclear shadowing of initial jet spectra, we shall use the centrality dependence of
charged particle pseudorapidity density, $dN_{\rm ch}/d\eta$, of the  ALICE data in 
Pb+Pb collisions to provide a more stringent constraint on the gluon shadowing parameter 
$s_g$ which will be then employed to investigate jet suppression. 
  
In the two-component HIJING model \cite{HIJING} for hadron production, nucleon-nucleon 
collision with transverse momentum $p_T$ transfer larger than a cut-off $p_0$ leads to 
jet production calculable by collinearly factorized pQCD model. 
Soft interactions with $p_T<p_0$ is characterized by an
effective cross section $\sigma_{\rm soft}$. In the HIJING 2.0 model \cite{HIJ2} the 
Duke-Owens parametrization \cite{Duke} of the parton distribution functions has been updated 
with the modern Gl\"uck-Reya-Vogt (GRV) parametrization \cite{GRV94}. Since the gluon 
distribution at small momentum fraction $x$ is much larger in GRV, instead of a fixed 
value for $p_0=2$ GeV/c and $\sigma_{\rm soft}=57$ mb
(as used in HIJING 1.0), an energy dependent cut-off for $p_0(\sqrt s)$ and  
$\sigma_{\rm soft}(\sqrt s)$ is used to fit experimental data on total and inelastic 
cross sections and hadron rapidity density in $p+p/{\bar p}$ collisions \cite{HIJ2}.

For the nuclear parton distribution function (PDF), HIJING employs the functional form
\begin{equation} \label{pdf}
f_a^A(x,Q^2) = A \: R_a^A(x,Q^2) \: f_a^N(x,Q^2) 
\end{equation}
where $f_a^N$ is the PDF in a nucleon. The nuclear modification factor
of quarks and gluons ($a\equiv q,g$) in HIJING 2.0 parametrization are \cite{HIJ2}
\begin{eqnarray} \label{nmod}
R_q^A(x,b) &=& 1 + 1.19\log^{1/6}\!\!A \ (x^3-1.2x^2+0.21x) \nonumber\\
&& - s_q(b) \: (A^{1/3}-1)^{0.6} \: (1-3.5x^{0.5}) \nonumber\\
&& \times \exp(-x^2/0.01) ,  \nonumber\\
R_g^A(x,b) &=& 1 + 1.19\log^{1/6}\!\!A \ (x^3-1.2x^2+0.21x) \nonumber\\
&& - s_g(b) \: (A^{1/3}-1)^{0.6} \: (1-1.5x^{0.35}) \nonumber\\
&& \times \exp(-x^2/0.004) . 
\end{eqnarray}
The impact parameter dependence of shadowing is taken as $s_a(b)=(5s_a/3)(1-b^2/R_a^2)$
that prohibits rapid rise of particle production with increasing centrality.
Here $R_A \sim A^{1/3}$ is the nuclear size and $s_q=0.1$ is fixed by data from
deep inelastic scatterings.  From comparison to the centrality dependence of charged particle 
pseudorapidity density per participant pair of nucleons, 
$(dN_{\rm ch}/d\eta)/(\langle N_{\rm part} \rangle/2)$ in Au+Au collisions at
$\sqrt{s_{NN}} =0.2$ TeV, the gluon shadowing parameter in HIJING 2.0 model has been 
constrained to $s_g=0.17-0.22$. Whereas a stronger constraint
of $s_g=0.20-0.23$ has been obtained from the reproduction of $dN_{\rm ch}/d\eta$ ALICE data 
for the most central (head-on) Pb+Pb collisions at $\sqrt{s_{NN}} =2.76$ TeV. Albeit, HIJING
ignores the final state interaction of particles, and such an estimate of $s_g$ is entirely
from initial state effects. We shall show the influence of final state parton energy 
loss \cite{GLV,WS,PalPratt} as well as hadronic rescatterings modify considerably 
the $dN_{\rm ch}/d\eta$ yield and thereby the magnitude of the initial state nuclear shadowing
$s_g$ for gluon distribution. In the present study we shall use the string melting version
of the AMPT where the hadrons from HIJING 2.0 are converted to their valence (anti)quarks
and parton recombination is employed for hadronization. The coalescence of dominant soft 
partons and also relatively large number of hard jets produced at LHC will thus contribute 
to the final charged hadron spectrum.
In the Lund string fragmentation function $f(z) \propto z^{-1}(1-z)^a\exp(-b m^2_T/z)$, where
$z$ is the light-cone momentum fraction of the generated hadrons with transverse mass
$m_T$, we employ the default HIJING values of $a=0.5$ and $b=0.9$ GeV$^{-2}$. 
Unless otherwise mentioned, at both RHIC and LHC energy we consider the strong coupling 
constant $\alpha_s=0.33$ and screening mass $\mu =3.226$ fm$^{-1}$ \cite{XuKo} that 
correspond to parton-parton elastic scattering cross section of $\sigma \approx 1.5$ mb 
in the parton cascade.

\begin{figure}[ht]
\centerline{\epsfig{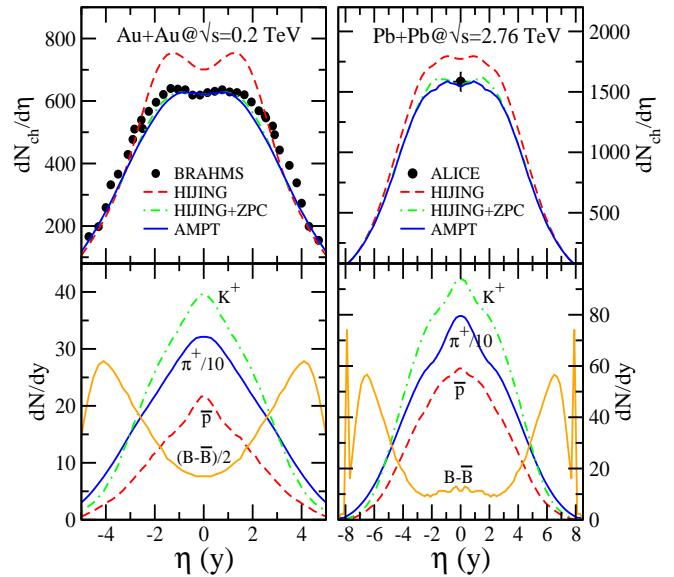}}
\vspace{-0.2cm}

\caption{Top panels: Pseudorapidity distribution for charged hadrons in Au+Au collision at RHIC 
energy of $\sqrt{s_{NN}}=0.2$ TeV and in Pb+Pb collision at LHC energy of $\sqrt{s_{NN}}=2.76$ TeV 
for ($0-5\%$) centrality. The AMPT model predictions are without any final state interactions as 
in HIJING (dashed line); with parton transport i.e. HIJING+ZPC (dashed-dotted line), and with 
further hadron transport as in AMPT (solid line). The solid circles are the measured values from 
the BRAHMS at RHIC \cite{BRAHMSnp} and ALICE at LHC \cite{ALICEnp}. 
Bottom panels: The rapidity distribution from AMPT for $K^+$, $\pi^+$, $\bar p$ and net baryons, 
$B-\bar B$, at the RHIC and LHC energies.}

\label{mulp}
\end{figure}

Figure \ref{mulp} shows the pseudorapidity distribution of charged hadrons in the $5\%$ 
most central collision in the AMPT model in Au+Au at $\sqrt{s_{NN}}=0.2$ TeV and 
Pb+Pb at $\sqrt{s_{NN}} = 2.76$ TeV. The results are with gluon shadowing parameter
of $s_g=0.15$ (at RHIC) and $s_g=0.17$ (at LHC) that are found to be in good agreement with 
the measured $dN_{\rm ch}/d\eta$ distribution from BRAHMS \cite{BRAHMS,BRAHMSnp} at 
$\sqrt{s_{NN}} = 0.2$ TeV, and the $dN_{\rm ch}/d\eta \: (|\eta|<0.5) = 1601\pm 60$ from ALICE 
\cite{ALICEnp} at $\sqrt{s_{NN}} = 2.76$ TeV. In absence of final state partonic and 
hadronic scatterings, which is basically the HIJING 2.0 model predicts
$dN_{\rm ch}/d\eta \: (|\eta|<0.5) = 706\pm 5$ and $1775 \pm 3$ at RHIC and LHC, respectively. 
In subsequent 
parton cascade (i.e. HIJING plus ZPC), energy dissipation and redistribution into the 
transverse flow via partonic scatterings lead to a reduction of charged particle multiplicity 
by surprisingly a similar amount of $\sim 15\%$ at both RHIC and LHC. Though the partonic 
density at LHC is about twice than at RHIC, this nearly equal suppression of yield after 
parton cascade reflects the interplay between hard and soft processes via a delicate balance 
between collective flow, gluon shadowing and jet multiplicity all of these are larger at LHC 
than at RHIC. Finally, subsequent hadronic scatterings (dubbed as AMPT) from the less dense phase 
leads to a smaller decrease of particle multiplicity.  
Fig. \ref{mulp} further shows that final state scatterings essentially smoothen out the dip at 
$\eta=0$ (due to Jacobian) in HIJING to a nearly flat $dN_{\rm ch}/d\eta$ distribution around 
mid-rapidity. Such a weak pseudorapidity-dependence in $dN_{\rm ch}/d\eta$ at $\eta \leq 2$ has 
also been observed in both the BRAHMS \cite{BRAHMSnp} and CMS data \cite{CMS,CMSnp}.

The rapidity distribution of pions, kaons, antiprotons and net baryons are displayed in 
Fig. \ref{mulp} at $\sqrt{s_{NN}}=0.2$ and 2.76 TeV. With more than an order of magnitude
increase in energy at LHC, 
the rapidity distribution of the produced hadrons becomes wider by $\sim 55\%$ and thereby
$dN_{\rm ch}/d\eta$ at midrapidity increases by $\sim 2.4$ compared to the top RHIC energy.
While the net-baryon density is found to decrease by $\sim 35\%$ from $\sqrt{s_{NN}}=0.2$ TeV
to 2.76 TeV, the antibaryon to baryon ratio at these RHIC (LHC) energies are found to be
$\bar p/p = 0.71 (0.88)$, $\bar\Lambda/\Lambda = 0.75 (0.95)$, $\bar\Xi/\Xi = 0.83 (0.99)$ and
$\bar\Omega/\Omega = 0.89 (1.00)$. The yield ratios from the AMPT at RHIC are consistent with 
the feed down corrected measured values \cite{PHENIX,BRAHMS} within the systematic errors. 
Enhanced meson production and slight decrease in the strangeness density at LHC result in the 
ratios at midrapidity of $p/\pi^+ = 0.091$ (0.088) and  $K^+/\pi^+ = 0.17$ (0.15) at the 
RHIC (LHC) energies considered here.

\begin{figure}[ht]
\centerline{\epsfig{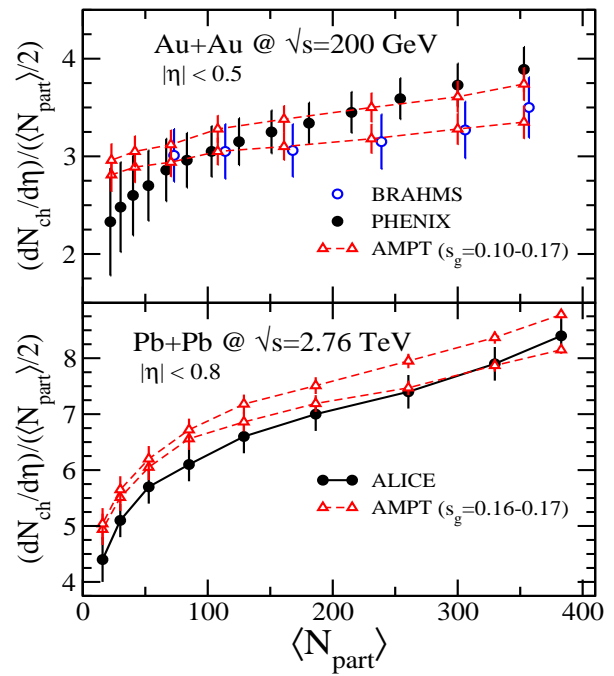}}
\vspace{-0.2cm}

\caption{Charged hadron multiplicity density $dN_{\rm ch}/d\eta$ at mid-rapidity per 
participant nucleon pair as a function of average number of participants 
$\langle N_{\rm part}\rangle$. The results are from AMPT calculations (triangles) 
obtained with gluon shadowing parameter $s_g=0.10-0.17$ in Au+Au collision at 
$\sqrt{s_{NN}}=0.2$ TeV (top panel) and with $s_g=0.16-0.17$ in Pb+Pb collision 
at $\sqrt{s_{NN}}=2.76$ TeV (bottom panel) as compared with the data (circles) from 
BRAHMS \cite{BRAHMSnp} and PHENIX \cite{PHENIXnp} at RHIC and ALICE \cite{ALICEnp} at LHC.}

\label{npart}
\end{figure}

In Fig. \ref{npart} we present the charged particle pseudorapidity density per participant pair, 
$(dN_{\rm ch}/d\eta)/(\langle N_{\rm part} \rangle/2)$, as a function of centrality of collision 
characterized by average number of participating nucleons $\langle N_{\rm part} \rangle$. 
The AMPT calculation are for Au+Au collisions at $\sqrt{s_{NN}} = 0.2$ TeV with a range of gluon
shadowing parameter $s_g=0.10-0.17$ and for Pb+Pb at $\sqrt{s_{NN}} = 2.76$ TeV with 
$s_g=0.16-0.17$.
With this choice of the gluon shadowing parameter, the centrality dependence of charged particle
multiplicity agrees well within the experimental uncertainty seen in the  BRAHMS \cite{BRAHMSnp} 
and PHENIX \cite{PHENIXnp} data at RHIC. Due to abundant jet and minijet production at LHC,
the ALICE multiplicity data for Pb+Pb collision is quite sensitive to nuclear distortions at
small $x$ and provides a much stringent constraint on the gluon shadowing of $s_g \simeq 0.17$.
It may be mentioned that the estimated values of $s_g$ in AMPT are consistently smaller
than in HIJING 2.0 model \cite{HIJ2} which underscores the importance of final state
interactions in precise estimation of the nuclear shadowing of partons that in turn
should also influence the hard observables.

The study of bulk hadron production when coupled with that for hadron spectra
provide crucial information of the parton-medium interactions where high-$p_T$ partons
suffer energy loss that are transported to produce soft hadrons. 
To quantify such a suppression of hadrons at high $p_T$ due to medium effects in heavy ion 
collisions, the nuclear modification factor
\begin{equation} \label{RAAeq}
R_{AA}(p_T) = \frac{d^2N^{AA}/d\eta \: dp_T} 
{\langle N_{\rm coll}\rangle \ d^2N^{pp}/d\eta \: dp_T} 
\end{equation}
is used which is the ratio of particle yield in heavy ions ($A+A$) to that in $p+p$ reference 
spectra, scaled by the total number of binary nucleon-nucleon ($NN$) collisions  
$\langle N_{\rm coll}\rangle = \langle T_{AA}\rangle \sigma^{NN}_{\rm inel}$. 
In absence of initial and final state nuclear medium effects $R_{AA}(p_T)=1$ by construction.
The nuclear thickness function $\langle T_{AA}\rangle$ and the inelastic $NN$ cross section
$\sigma^{NN}_{\rm inel}$ are calculated within the HIJING 2.0 model that uses Glauber Monte
Carlo simulation for distribution of initial nucleons with a Woods-Saxon nuclear density.
The energy dependent soft interaction cross section $\sigma_{\rm soft}(\sqrt s)$ in 
HIJING 2.0 enforces $\sigma^{NN}_{\rm inel}$ to be about 42 and 64 mb at 
$\sqrt{s_{NN}} = 0.2$ and 2.76 TeV, respectively. However, at low $p_T$ regime dominated 
by soft particle production, the scaling by the number of nucleons suffering at least 
one inelastic collision, i.e $N_{\rm part}$, is more appropriate.

\begin{figure}[ht]
\centerline{\epsfig{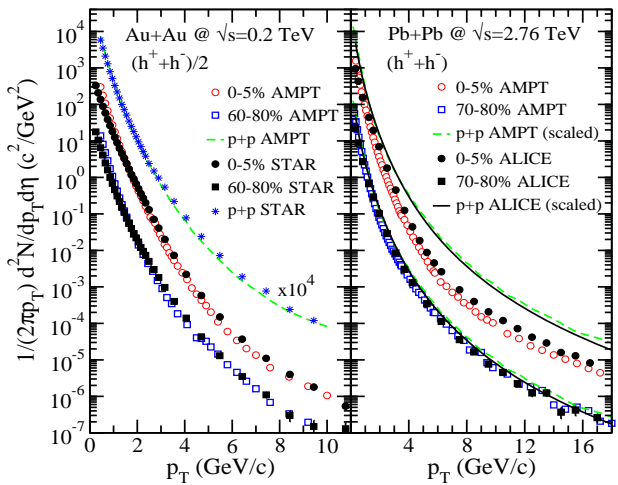}}
\vspace{-0.2cm}

\caption{Invariant hadron production spectrum in $p+p$ and Au+Au collision at $\sqrt{s_{NN}}=0.2$ TeV 
(left panel) and in $p+p$ and Pb+Pb collision at $\sqrt{s_{NN}}=2.76$ TeV (right panel).
The results are from AMPT calculations in $p+p$ (dashed lines) and heavy ion ($A+A$) 
(open symbols) collisions with gluon shadowing parameter $s_g=0.15$ (0.17) at RHIC (LHC). 
The measured spectrum are for Au+Au (solid symbols) and $p+p$ non-single-diffractive 
interaction (star) by STAR \cite{STARsup} at RHIC and for Pb+Pb (solid symbols)  
by ALICE \cite{ALICEsup} at LHC. The $p+p$ reference spectrum at $\sqrt{s_{NN}}=2.76$ TeV 
(solid lines) is the ALICE interpolation normalized to LO pQCD which is shown as scaled 
by average number of binary collisions, $\langle N_{\rm coll}\rangle$, corresponding to 
the centrality classes.}

\label{spectra}
\end{figure}

Figure \ref{spectra} shows the inclusive charged hadron $p_T$ spectra at midrapidity
in the AMPT for $p+p$ collisions and for central ($0-5\%$) and peripheral Au+Au
collisions at $\sqrt{s_{NN}} = 200$ GeV (left panel) and in Pb+Pb collisions at 
$\sqrt{s_{NN}}=2760$ GeV (right panel).
The results are for initial parton distribution with gluon shadowing $s_g= 0.15$ (0.17) at 
RHIC (LHC) energies that have been fixed from the centrality dependence of $N_{\rm ch}$ data. 
In $p+p$ collisions, the $p_T$ spectra from the model exhibit the LO pQCD based power law behavior 
at $p_T > 5$ GeV/c which is in overall good agreement with the STAR data \cite{STARsup}. 
At $\sqrt{s_{NN}}=2.76$ TeV, we however find the calculated yield from $p+p$ 
overpredicts at $p_T \gtrsim 6$ GeV/c that obtained by ALICE \cite{ALICEsup} 
from interpolation of $p\bar p$ spectrum measurements at $\sqrt{s_{NN}}=0.9$ and 7 TeV to 
estimate the suppression $R_{AA}$. For peripheral heavy ion collisions the AMPT spectra are 
consistent with that measured at both RHIC and LHC energies. On the other hand, the $p_T$
distributions for central collision
show marked deviation from power law function and are clearly suppressed especially at
moderate $p_T=4-11$ GeV/c due to medium modification. Though  
the AMPT spectra from central collisions describes the RHIC data quite well, 
it is however much softer than the ALICE data at $p_T > 2$ GeV/c. This possibly stems from 
enhanced energy loss of partons in a much denser medium that is generated from melting 
of strings to their valence quarks and antiquarks in the QGP medium.

\begin{figure}[ht]
\centerline{\epsfig{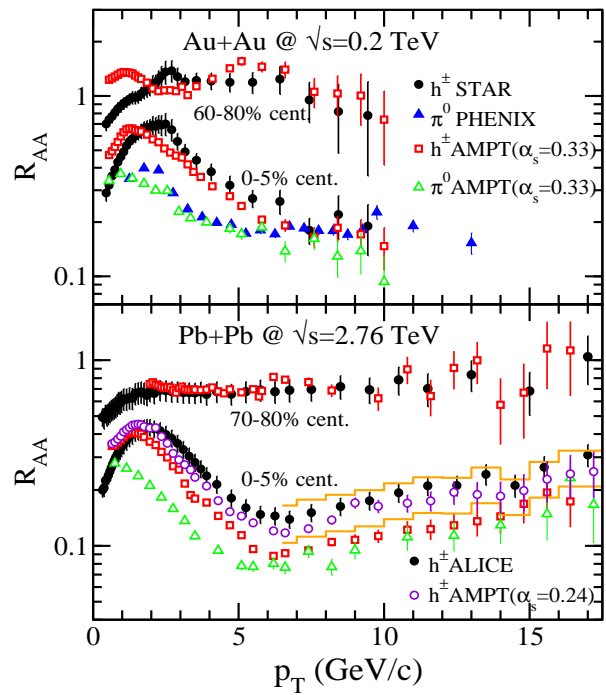}}
\vspace{-0.2cm}

\caption{Nuclear modification factor $R_{AA}$ for charged hadrons and neutral pions 
as a function of $p_T$ in central and peripheral Au+Au collisions at $\sqrt{s_{NN}}=0.2$ TeV 
(top panel) and Pb+Pb collisions at $\sqrt{s_{NN}}=2.76$ TeV (bottom panel). The AMPT model 
predictions are compared to the data from STAR \cite{STARsup} and PHENIX \cite{PHENIXsup} 
at RHIC and from ALICE \cite{ALICEsup} at LHC. The AMPT results are with strong coupling
constant $\alpha_s=0.33$ at RHIC and LHC and with $\alpha_s=0.24$ for central 
collisions at LHC. The histograms is the systematic error band due to different interpolation 
procedure used in earlier estimates by ALICE for the baseline $p+p$ spectra.}

\label{RAA}
\end{figure}

The nuclear modification factor $R_{AA}$ for charged hadrons is shown in Fig. \ref{RAA} 
for central and peripheral Au+Au collision at RHIC (top panel) and in Pb+Pb collisions 
at LHC (bottom panel).
For central collisions at both energies, $R_{AA}(p_T)$ is less than unity which implies
appreciable suppression of charged hadrons relative to $NN$ reference. The model
calculations, with nuclear shadowing parameter $s_g=0.15$ constrained from
$dN_{\rm ch}/d\eta$ data in Au+Au collisions, describes the magnitude and pattern of
the RHIC suppression data \cite{STARsup}. It is seen that $R_{AA}$ increases gradually with 
$p_T$ reaches a maximum of $R_{AA} \simeq 0.7$ at $p_T \simeq 1.8$ GeV/c, then it decreases
with further increase of $p_T$ and saturates thereafter to about 0.2 at $p_T \gtrsim 7$ GeV/c.
The success of AMPT at $\sqrt{s_{NN}}=0.2$ TeV thus suggests that the initial state
shadowing of pQCD jet spectra, the final state scattering and the parton energy loss
is consistent with the formation and evolution of the medium at RHIC energy.

At $70-80\%$ centrality Pb+Pb collisions at $\sqrt{s_{NN}}=2.76$ TeV, the $R_{AA}$ for
charged hadrons is nearly 
constant at about 0.7 over a large $p_T$ range as seen in both the ALICE data \cite{ALICEsup} 
and AMPT model calculations. At this peripheral collision, the QGP even if
formed, should have a small volume and short lifetime. In central Pb+Pb collisions
at LHC, the rise and fall pattern exhibited by $R_{AA}$ up to $p_T \sim 6$ GeV/c is
similar to RHIC. However, as evident from ALICE data, the suppression of charged hadrons 
at low $p_T$ is somewhat larger, and $R_{AA}$ reaches a minimum of 0.14 around 6-7 GeV/c.
The earlier estimates with large errors as shown by histogram is due to interpolation
procedure used by ALICE to obtain the baseline $p+p$ spectrum. With the recently measured
spectrum in $p+p$ collision at $\sqrt{s_{NN}} = 2.76$ TeV \cite{ALICE}, the measured $R_{AA}$ 
drops but remains well within the systematic error bands which is also consistent with the 
CMS data \cite{CMSsup}. In contrast to ALICE data, the AMPT calculations show even more 
pronounced suppression at $p_T>2$ GeV/c due to 
significant quenching of the hard-scattered partons. Within the coalescence mechanism
for hadronization in AMPT, though the peak positions and the subsequent decreasing pattern of
$R_{AA}$ are similar to the measured RHIC and LHC data, the minimum is found to be at 0.09
at $p_T \sim 6$ GeV/c. The subsequent rise of $R_{AA}$ (compared to nearly constant
value at RHIC) essentially stems from harder unquenched pQCD jet spectra at LHC and
found to have similar slope as in the data. 

In Fig. \ref{RAA} we also show the $R_{AA}$ for neutral pions for central collisions.
As seen in charged hadrons, the $R_{AA}$ for $\pi^0$ exhibit a similar but a gradual 
rise and fall pattern at intermediate $p_T$ ($1.8 <p_T < 5$ GeV/c) and thereafter saturates 
(rises) with increasing $p_T$ at RHIC (LHC) energies. Both the calculation and PHENIX
data show that relative to charged hadrons, the $\pi^0$s are more suppressed 
by as much as $\sim 45\%$ at the intermediate $p_T$. However, at $p_T \gtrsim 5$ GeV/c the
magnitude of suppression are nearly same for neutral pion and charged hadrons.  
The larger $R_{AA}$ for charged hadrons compared to neutral pions can be explained
as due to large baryonic (protons and antiprotons) yield produced from parton
coalescence used for hadronization \cite{Greco,Fries}. In fact we find the invariant
yield of pions and protons become comparable at $p_T \sim 2-4$ GeV/c. 
At $p_T \gtrsim 6$ GeV/c, as pions are the most abundant particles and moreover
the parton spectra become gradually flatter with increasing $p_T$, coalescence
of hard partons is seen in AMPT to predict in nearly identical suppression 
$R_{AA}$ for pions and hadrons.  

The significant suppression in AMPT much below than the ALICE data suggests that the medium 
with more than a factor of two larger parton density than RHIC is in fact more transparent than 
expected. Attempt to increase $R_{AA}$ at high $p_T$ by decreasing the shadowing $s_g$ 
only result in an enhanced bulk (soft) hadron production and thus disagree with the centrality 
dependence of $dN_{\rm ch}/d\eta$ data shown in Fig. \ref{npart}. In fact, the WDGH jet energy 
loss model \cite{Horowitz} that has been constrained to fit the RHIC suppression data severely 
underpredicts the central $R_{AA}$ value of ALICE. 

On the other hand, we note that the above suppression was calculated in the AMPT with same values 
of QCD coupling constant $\alpha_s=0.33$ and screening mass $\mu =3.226$ fm$^{-1}$ at both RHIC 
and LHC. Perturbatively, the screening mass depends on temperature as $\mu = gT$ with 
$g=\sqrt{4\pi\alpha_s}$ \cite{Blaizot}. The parton-parton elastic scattering cross section
used in AMPT then reduces to $\sigma \approx 9\pi\alpha_s^2/(2\mu^2) \approx 9\alpha_s/(8T^2)$.
Since hydrodynamic model analysis of RHIC/LHC data indicate \cite{Song08} only about 
$10\%$ viscous entropy production, the initial entropy density $s_i$ can be approximated to 
final particle multiplicity by the scaling \cite{Gyulassy}
\begin{eqnarray} \label{entr}
s_i \approx \frac{1}{\tau_i A_\perp} \frac{dS}{dy} 
\approx \frac{1}{\tau_i A_\perp} 7.85 \frac{dN_{\rm ch}}{dy} ,
\end{eqnarray}
where $A_\perp$ is the transverse area of the collision zone. The proportionality constant 
for entropy rapidity density, $dS/dy$, to $dN_{\rm ch}/dy$ conversion was taken from 
Refs. \cite{Gyulassy,Kolb,PPs}. For a QGP characterized by massless gas of light 
quarks and antiquarks, $s_i \approx 4 \epsilon_i /(3T_i)$ with energy density 
$\epsilon_i \approx (21/30)\pi^2 T_i^4$. This allows to estimate the initial temperature 
$T_i$ and thereby the parton scattering cross section $\sigma$ from 
the measured particle yield. For the $5\%$ most central Au+Au and Pb+Pb collisions at   
$\sqrt{s_{NN}} = 200$ and 2076 GeV, the measured $dN_{\rm ch}/dy \approx 687$ \cite{PHENIXnp} 
and 1601 \cite{ALICEnp} result in $T_i \approx 320$ and 436 MeV respectively, at a proper 
time $\tau_i=1$ fm/c. With the above choice of $\alpha_s=0.33$, the estimated
$\sigma \approx 9\alpha_s/(8T^2) \approx 1.4$ mb at RHIC is incidentally close to the 
value employed in AMPT that reproduces the RHIC suppression data shown in Fig. \ref{RAA}. 
In contrast, the higher temperature $T_i$ at LHC enforces a much smaller 
$\sigma \approx 0.76$ mb. Alternatively, if the screening mass remains constant at 
$\mu =3.226$ fm$^{-1}$ from RHIC to LHC, such a small $\sigma$ is then consistent with 
$\alpha_s \approx 0.24$ at LHC. With this reduced $\alpha_s$, we show in Fig. \ref{RAA} 
(open circles) the AMPT results 
of $R_{AA}$ of charged hadrons in central Pb+Pb collisions at $\sqrt{s_{NN}} = 2.76$ TeV.
The good agreement with the ALICE suppression data is a clear indication of thermal 
suppression of the QCD coupling constant due to higher temperature at LHC compared to 
that at RHIC. It may however be mentioned that instead of an average value, the strong 
coupling could have a temperature dependence $\alpha_s(T)$ during the plasma evolution 
\cite{Hirano}. Further, the AMPT model calculations invoke purely elastic collisional 
energy loss, the effects of inelastic scatterings via medium-induced radiative parton energy 
loss \cite{GLV,WS,BAMPS,Horowitz} could still pose a serious theoretical challenge to 
understand the underlying energy loss mechanism especially at the LHC energy regime.

In summary, we study the nuclear medium effects on hadron production over a wide 
range of $p_T$ in Pb+Pb collisions at the LHC energy $\sqrt{s_{NN}} = 2076$ GeV.
For this purpose we use the AMPT model which is updated to include the HIJING 2.0 
version for initial conditions for parton distribution. We find final-state parton 
scatterings reduce significantly the hadron multiplicity at midrapidity that enforces 
smaller gluon shadowing for agreement with the ALICE data for charge particle yield
at various centralities. With such a constrained parton shadowing, we find
that parton energy loss in AMPT describes quite well the magnitude and 
suppression pattern of hadrons in both central and peripheral Au+Au collisions 
at the RHIC energy $\sqrt{s_{NN}} = 200$ GeV. With the same strong coupling constant
$\alpha_s=0.33$, the model however predicts larger jet quenching relative to ALICE data 
for central Pb+Pb collisions at LHC. A reduction of $\alpha_s$ by $\sim 30\%$ in the higher
temperature plasma formed at LHC was found to describe the measured suppression.

\bigskip
SP acknowledges support from Helmholtz International Center for FAIR and kind hospitality 
at the Frankfurt Institute for Advanced Studies where part of the work was developed.


\begin{thebibliography}{99}

\bibitem{BRAHMS} I. Arsene et. al., Nucl. Phys. A 757 (2005) 1.
\bibitem{PHOBOS} B.B. Back et. al., Nucl. Phys. A 757 (2005) 28.
\bibitem{STAR} J. Adams et. al., Nucl. Phys. A 757 (2005) 102.
\bibitem{PHENIX} K. Adcox et. al., Nucl. Phys. A 757 (2005) 184.
\bibitem{ALICE} J. Schukraft for the ALICE Collaboration, J. Phys. G  (2011) 124003.
\bibitem{CMS} B. Wyslouch for the CMS Collaboration, J. Phys. G 38 (2011) 124005.
\bibitem{DIS} M. Gyulassy and L. McLerran, Nucl. Phys. A 750 (2005) 30.
\bibitem{Shuryak} E.V. Shuryak, Nucl. Phys. A 750 (2005) 64.
\bibitem{STARsup} J. Adams et. al., Phys. Rev. Lett. 91 (2003) 172302.
\bibitem{PHENIXsup} S.S. Adler et. al., Phys. Rev. C 69 (2004) 034910.
\bibitem{GLV} M. Gyulassy, P. Levai, and I. Vitev, Phys. Rev. Lett. 85 (2000) 5535.
\bibitem{WS} M. Gyulassy, I. Vitev, X.N. Wang, and B.W. Zhang, in
{\em Quark Gluon Plasma}, ed. by R.C. Hwa and X.N. Wang (World Scientific,
Singapore, 2004); A. Kovner and U.A. Wiedemann in {\em Quark Gluon Plasma},
ed. by R.C. Hwa and X.N. Wang (World Scientific, Singapore, 2004).
\bibitem{PalPratt} S. Pal and S. Pratt, Phys. Lett. B 574 (2003) 21; S. Pal
Phys. Rev. C 80 (2009) 041901(R).
\bibitem{ALICEnp} K. Aamodt et. al.,  Phys. Rev. Lett. 106 (2011) 032301.
\bibitem{CMSnp} CMS Collaboration, arXiv:1107.4800.
\bibitem{ALICEsup} K. Aamodt et. al., Phys. Lett. B 696 (2011) 30.
\bibitem{CMSsup} A.S. Yoon for CMS Collaboration, J. Phys. G 38 (2011) 124116.
\bibitem{Paul} P. Romatschke and U. Romatschke, Phys. Rev. Lett. 99 (2007) 172301.
\bibitem{Song} H. Song and U. Heinz, Phys. Lett. B 658 (2008) 279.
\bibitem{Chen} X.-F. Chen, T. Hirano, E. Wang, X.-N. Wang, and H. Zhang, 
Phys. Rev. C 84 (2011) 034902.
\bibitem{UrQMD} S. Scherer et al. Prog. Part. Nucl. Phys. 42 (1999) 279.
\bibitem{AMPT} Z.W. Lin, C.M. Ko, B.-A. Li, B. Zhang, and S. Pal,
Phys. Rev. C 72 (2005) 064901.
\bibitem{BAMPS} Z. Xu and C. Greiner, Phys. Rev. C 71 (2005) 064901; 
{\em ibid} 76 (2007) 024911.
\bibitem{Petersen} H. Petersen, J. Steinheimer, G. Burau, M. Bleicher and H. St\"ocker, 
Phys. Rev. C 78 (2008) 044901.
\bibitem{HIJ2} W.-T. Deng, X.-N. Wang, and R. Xu,  
Phys. Rev. C 83 (2011) 014915; Phys. Lett. B 701 (2011) 133.
\bibitem{Li} S. Li and X.N. Wang, Phys. Lett. B 527 (2002) 85.
\bibitem{HIJING} X.N. Wang and M. Gyulassy, Phys. Rev. D 44 (1991) 3501.
\bibitem{Duke} D.W. Duke and J.F. Owens, Phys. Rev. D 30 (1984) 49.
\bibitem{GRV94} M. Gl\"uck, E. Reya, and W. Vogt, Z. Phys. C 67 (1995) 433.
\bibitem{XuKo} J. Xu and C.M. Ko, Phys. Rev. C 83 (2011) 034904.
\bibitem{BRAHMSnp} I.G. Bearden, et. al., Phys. Rev. Lett. 88 (2002) 202301.
\bibitem{PHENIXnp} S.S. Adler, et. al., Phys. Rev. C 71 (2005) 034908.
\bibitem{Greco} V. Greco, C.M. Ko, and P. Levai, Phys. Rev. Lett. 90 (2003) 202302.
\bibitem{Fries} R.J. Fries, B. Muller, C. Nonaka, and S.A. Bass, 
Phys. Rev. Lett. 90 (2003) 202303.
\bibitem{Horowitz} W.A. Horowitz and M. Gyulassy, Nucl. Phys. A 872 (2011) 265.
\bibitem{Blaizot} J.P. Blaizot and E. Iancu, Phys. Rep. 359 (2002) 355.
\bibitem{Song08} H. Song and U.W. Heinz, Phys. Rev. C 78 (2008) 024902.
\bibitem{Gyulassy} M. Gyulassy and T. Matsui, Phys. Rev. D 29 (1984) 419.
\bibitem{Kolb} P.F Kolb, U.W. Heinz, P. Huovinen, K.J. Eskola, and
K. Tuominen, Nucl. Phys. A 696 (2001) 197.
\bibitem{PPs} S. Pal and S. Pratt, Phys. Lett. B 578 (2004) 310.
\bibitem{Hirano} T. Hirano and M. Gyulassy, Nucl. Phys A 769 (2006) 71.


\end{thebibliography}
\end{document}